\documentclass[12pt]{article}
\usepackage{graphicx}
\usepackage{lscape}
\usepackage{citesort}
\usepackage{amssymb}
\usepackage{appendix}
\usepackage{multirow}

\begin{document}
\def\qq{\langle \bar q q \rangle}
\def\uu{\langle \bar u u \rangle}
\def\dd{\langle \bar d d \rangle}
\def\sp{\langle \bar s s \rangle}
\def\GG{\langle g_s^2 G^2 \rangle}
\def\Tr{\mbox{Tr}}
\def\figt#1#2#3{
        \begin{figure}
        $\left. \right.$
        \vspace*{-2cm}
        \begin{center}
        \includegraphics[width=10cm]{#1}
        \end{center}
        \vspace*{-0.2cm}
        \caption{#3}
        \label{#2}
        \end{figure}
    }

\def\figb#1#2#3{
        \begin{figure}
        $\left. \right.$
        \vspace*{-1cm}
        \begin{center}
        \includegraphics[width=10cm]{#1}
        \end{center}
        \vspace*{-0.2cm}
        \caption{#3}
        \label{#2}
        \end{figure}
                }

\def\ds{\displaystyle}
\def\beq{\begin{equation}}
\def\eeq{\end{equation}}
\def\bea{\begin{eqnarray}}
\def\eea{\end{eqnarray}}
\def\beeq{\begin{eqnarray}}
\def\eeeq{\end{eqnarray}}
\def\ve{\vert}
\def\vel{\left|}
\def\ver{\right|}
\def\nnb{\nonumber}
\def\ga{\left(}
\def\dr{\right)}
\def\aga{\left\{}
\def\adr{\right\}}
\def\lla{\left<}
\def\rra{\right>}
\def\rar{\rightarrow}
\def\lrar{\leftrightarrow}
\def\nnb{\nonumber}
\def\la{\langle}
\def\ra{\rangle}
\def\ba{\begin{array}}
\def\ea{\end{array}}
\def\tr{\mbox{Tr}}
\def\ssp{{\Sigma^{*+}}}
\def\sso{{\Sigma^{*0}}}
\def\ssm{{\Sigma^{*-}}}
\def\xis0{{\Xi^{*0}}}
\def\xism{{\Xi^{*-}}}
\def\qs{\la \bar s s \ra}
\def\qu{\la \bar u u \ra}
\def\qd{\la \bar d d \ra}
\def\qq{\la \bar q q \ra}
\def\gGgG{\la g^2 G^2 \ra}
\def\q{\gamma_5 \not\!q}
\def\x{\gamma_5 \not\!x}
\def\g5{\gamma_5}
\def\sb{S_Q^{cf}}
\def\sd{S_d^{be}}
\def\su{S_u^{ad}}
\def\sbp{{S}_Q^{'cf}}
\def\sdp{{S}_d^{'be}}
\def\sup{{S}_u^{'ad}}
\def\ssp{{S}_s^{'??}}

\def\sig{\sigma_{\mu \nu} \gamma_5 p^\mu q^\nu}
\def\fo{f_0(\frac{s_0}{M^2})}
\def\ffi{f_1(\frac{s_0}{M^2})}
\def\fii{f_2(\frac{s_0}{M^2})}
\def\O{{\cal O}}
\def\sl{{\Sigma^0 \Lambda}}
\def\es{\!\!\! &=& \!\!\!}
\def\ap{\!\!\! &\approx& \!\!\!}
\def\ar{&+& \!\!\!}
\def\ek{&-& \!\!\!}
\def\kek{\!\!\!&-& \!\!\!}
\def\cp{&\times& \!\!\!}
\def\se{\!\!\! &\simeq& \!\!\!}
\def\eqv{&\equiv& \!\!\!}
\def\kpm{&\pm& \!\!\!}
\def\kmp{&\mp& \!\!\!}
\def\mcdot{\!\cdot\!}
\def\erar{&\rightarrow&}


\def\simlt{\stackrel{<}{{}_\sim}}
\def\simgt{\stackrel{>}{{}_\sim}}


\renewcommand{\textfraction}{0.2}    
\renewcommand{\topfraction}{0.8}

\renewcommand{\bottomfraction}{0.4}
\renewcommand{\floatpagefraction}{0.8}
\newcommand\mysection{\setcounter{equation}{0}\section}

\def\baeq{\begin{appeq}}     \def\eaeq{\end{appeq}}
\def\baeeq{\begin{appeeq}}   \def\eaeeq{\end{appeeq}}
\newenvironment{appeq}{\beq}{\eeq}
\newenvironment{appeeq}{\beeq}{\eeeq}
\def\bAPP#1#2{
 \markright{APPENDIX #1}
 \addcontentsline{toc}{section}{Appendix #1: #2}
 \medskip
 \medskip
 \begin{center}      {\bf\LARGE Appendix #1 :}{\quad\Large\bf #2}
\end{center}
 \renewcommand{\thesection}{#1.\arabic{section}}
\setcounter{equation}{0}
        \renewcommand{\thehran}{#1.\arabic{hran}}
\renewenvironment{appeq}
  {  \renewcommand{\theequation}{#1.\arabic{equation}}
     \beq
  }{\eeq}
\renewenvironment{appeeq}
  {  \renewcommand{\theequation}{#1.\arabic{equation}}
     \beeq
  }{\eeeq}
\nopagebreak \noindent}

\def\eAPP{\renewcommand{\thehran}{\thesection.\arabic{hran}}}

\renewcommand{\theequation}{\arabic{equation}}
\newcounter{hran}
\renewcommand{\thehran}{\thesection.\arabic{hran}}

\def\bmini{\setcounter{hran}{\value{equation}}
\refstepcounter{hran}\setcounter{equation}{0}
\renewcommand{\theequation}{\thehran\alph{equation}}\begin{eqnarray}}
\def\bminiG#1{\setcounter{hran}{\value{equation}}
\refstepcounter{hran}\setcounter{equation}{-1}
\renewcommand{\theequation}{\thehran\alph{equation}}
\refstepcounter{equation}\label{#1}\begin{eqnarray}}


\newskip\humongous \humongous=0pt plus 1000pt minus 1000pt
\def\caja{\mathsurround=0pt}


\title{
         {\Large
                 {\bf
Strong transitions of decuplet  to octet baryons and pseudoscalar mesons
                 }
         }
      }

\author{\vspace{1cm}\\
{\small T. M. Aliev$^a$ \thanks {e-mail:
taliev@metu.edu.tr}~\footnote{permanent address: Institute of
Physics, Baku, Azerbaijan}\,\,, K. Azizi$^b$ \thanks {e-mail:
kazizi@dogus.edu.tr}\,\,, M. Savc{\i}$^a$ \thanks
{e-mail: savci@metu.edu.tr}} \\
{\small $^a$ Physics Department, Middle East Technical University,
06531 Ankara, Turkey} \\
{\small $^b$ Physics Division, Faculty of Arts and
Sciences, Do\u gu\c s University} \\
{\small Ac{\i}badem-Kad{\i}k\"oy, 34722 Istanbul, Turkey} }

\date{}

\begin{titlepage}
\maketitle
\thispagestyle{empty}

\begin{abstract}
The strong coupling constants of light pseudoscalar $\pi$, $K$ and
$\eta$ mesons with decuplet--octet baryons are studied within light
cone QCD sum rules, where $SU(3)_f$ symmetry breaking effects are
taken into account. It is shown that all coupling constants under
the consideration are described by only one universal function even
if $SU(3)_f$ symmetry breaking effects are switched  into the game.
\end{abstract}

~~~PACS number(s): 11.55.Hx, 13.75.Gx, 13.75.Jz
\end{titlepage}

\section{Introduction}
During recent years, intense studies have been made in the pion and
kaon photo and electroproduction off the nucleon. Several exiting
experimental programs exploiting these reactions have already been
performed at electron beam facilities such as MIT--Bates, MAMI and
Jefferson Laboratory. One main goal of these experiments is
determination of the coupling constants of pion and kaon with
baryons. Calculation of the coupling constants of pseudoscalar
mesons with hadrons in the framework of QCD, is also very important
for understanding the dynamics of pion and kaon photo and
electroproduction reaction off the nucleon.

These coupling constants belong to the low energy sector of QCD, which is far from perturbative regime. Therefore, for calculation of these coupling constants some non-perturbative methods are needed.
 QCD sum rules method \cite{R10501} is one of the most
promising and predictive one among all existing non-perturbing
methods in studying the properties of hadrons. In this work, we
calculate the coupling constants of pseudoscalar mesons with
decuplet--octet baryons within the light cone (LCSR) method (for
more about this method, see \cite{R10502}). In light cone QCD sum rules, the operator product expansion (OPE) is carried out near the light cone, $x^2\simeq0$,
 instead of the short distance, $x\simeq0$ in traditional ones. In this approach, the OPE is also carried out over twist rather than dimension of operators in traditional sum rules. The main ingredient of LCSR are distribution amplitudes (DA's) which appear in matrix elements of nonlocal operators between the vacuum and the one-particle states.

 Present work is an
extension of our previous works, where coupling constants of
pseudoscalar and vector mesons with octet baryons
\cite{R10503,R10504}, pseudoscalar mesons with decuplet baryons
\cite{R10505} and vector mesons with decuplet--octet baryons
\cite{R10506} are calculated. Here, using the DA's of the pseudoscalar mesons, we calculate the strong coupling constants of light pseudoscalar $\pi$, $K$ and
$\eta$ mesons with decuplet--octet baryons in the framework of the  light
cone QCD sum rules both in full theory and when the  $SU(3)_f$ symmetry breaking effects are taken into account.  We would like especially to note that the main advantage of the approach
presented in this work  is that it takes into account the $SU(3)_f$
symmetry violation effects automatically.

The  paper is organized as follows. In section 2, the strong
coupling constants of pseudoscalar mesons with decuplet--octet
baryons are calculated within LCSR method. In this section, we also
obtain the relations between correlation functions describing
various coupling constants. It is also shown that all coupling constants under
the consideration are described by only one universal function even
if $SU(3)_f$ symmetry breaking effects are considered.  In section 3, we present our numerical
analysis of the coupling constants of pseudoscalar mesons with
decuplet--octet baryons. This section also includes  comparison of  our predictions on the coupling constants  with the existing experimental data.

\section{Sum rules for the coupling constants of the pseudoscalar mesons
with decuplet--octet baryons}

In this section, we obtain LCSR for the coupling constants of
pseudoscalar mesons with decuplet--octet baryons. Before calculating
these coupling constants within LCSR, it should be remembered that
the within $SU(3)_f$ symmetry, the coupling of pseudoscalar mesons
with decuplet--octet baryons is described by the single coupling
constant whose interaction Lagrangian is given by \bea
\label{e10501} {\cal L}_{\rm int} = g_{\cal DOP} \varepsilon_{ijk}
\bar{\cal O}_\ell^j ({\cal D}^{mk\ell})_\mu \partial^\mu {\cal
P}_m^i + \mbox{\rm h.c.}~, \eea where ${\cal O}$, ${\cal D}$ and
${\cal P}$ correspond to octet, decuplet baryons and pseudoscalar
mesons, respectively, and $g_{\cal DOP}$ is the coupling constant of
the pseudoscalar mesons with decuplet--octet baryons. After this
preliminary remark, we can proceed to derive the sum rules for the
strong coupling constants of the pseudoscalar mesons with
decuplet--octet baryons. For this purpose, we consider the
correlation function \bea \label{e10502} \Pi_\mu = i \int d^4x
e^{ipx} \lla {\cal P}(q) \vel {\cal T}\left\{ \eta(x)
\bar{\eta}_\mu(0) \right\} \ver 0 \rra~, \eea where ${\cal P}(q)$ is
the pseudoscalar meson with momentum $q$, $\eta_\mu$ and $\eta$ are
the interpolating currents for decuplet and octet baryons,
respectively, and ${\cal T}$ is the time ordering operator.  The sum
rules for the coupling constants can be obtained by calculating the
correlation function in terms of hadrons and also in deep Euclidean
region, where $-p^2 \rar \infty$ and $-(p+q)^2 \rar \infty$, in terms of quark and gluon degrees of
freedom, and then equating these expressions using the dispersion
relation. Note that in short-distance version of sum rules, where the operator product expansion is performed at
$x\simeq0$, the similar correlation function have been widely used in calculation of
 the pion-nucleon coupling constants in many works \cite{Doi,Shiomi,Birse,Kim}.

 It follows from Eq. (\ref{e10502}) that, in calculating
the phenomenological and theoretical parts we need expressions of
the interpolating currents for decuplet and octet baryons. The
general form of the interpolating currents of the octets and
decuplets are as follows \cite{R10507,R10508,R10509}: \bea
\label{e10503} \eta \es A \varepsilon^{abc} \left\{ (q_1^{aT} C
q_2^b) \gamma_5 q_3^c - (q_2^{aT} C q_3^b) \gamma_5 q_1^c + \beta
(q_1^{aT} C \gamma_5 q_2^b) q_3^c
-\beta (q_2^{aT} C\gamma_5 q_3^b)q_1^c\right\}~, \\
\label{e10504} \eta_\mu \es A^\prime \varepsilon^{abc} \left\{
(q_1^{aT} C \gamma_\mu q_2^b) q_3^c + (q_2^{aT} C \gamma_\mu q_3^b)
q_1^c + (q_3^{aT} C \gamma_5 q_1^b) q_2^c\right\}~, \eea where
$a,b,c$ are the color indices, $\beta$ is an arbitrary parameter,
$C$ is the charge conjugation operator. The values of normalization
constants $A$ and $A'$ and  the $q_1$, $q_2$ and $q_3$ quarks for
each octet and decuplet baryon are represented in Tables 1 and 2,
respectively.
\begin{table}[h]

\renewcommand{\arraystretch}{1.3}
\addtolength{\arraycolsep}{-0.5pt}
\small
$$
\begin{array}{|l|c|c|c|c|}
\hline \hline
 & A & q_1 & q_2 & q_3 \\  \hline
 \Sigma^0 & -\sqrt{1/2}  & u & s & d  \\
 \Sigma^+ &        1/2   & u & s & u  \\
 \Sigma^- &        1/2   & d & s & d  \\
 p        &       -1/2   & u & d & u  \\
 n        &       -1/2   & d & u & d  \\
 \Xi^0    &        1/2   & s & u & s  \\
 \Xi^-    &        1/2   & s & d & s  \\
\hline \hline
\end{array}
$$
\caption{The values of $A$ and the quark flavors $q_1$, $q_2$ and
$q_3$ for octet baryons.}
\renewcommand{\arraystretch}{1}
\addtolength{\arraycolsep}{-1.0pt}

\end{table}
\begin{table}[h]

\renewcommand{\arraystretch}{1.3}
\addtolength{\arraycolsep}{-0.5pt}
\small
$$
\begin{array}{|l|c|c|c|c|}
\hline \hline
 & A^\prime & q_1 & q_2 & q_3 \\  \hline
 \Sigma^{\ast 0} & \sqrt{2/3}  & u & d & s  \\
 \Sigma^{\ast +} & \sqrt{1/3}  & u & u & s  \\
 \Sigma^{\ast -} & \sqrt{1/3}  & d & d & s  \\
 \Delta^{++}     &       1/3   & u & u & u  \\
 \Delta^{+}      & \sqrt{1/3}  & u & u & d  \\
 \Delta^{0}      & \sqrt{1/3}  & d & d & u  \\
 \Delta^{-}      &       1/3   & d & d & d  \\
 \Xi^{\ast 0}    & \sqrt{1/3}  & s & s & u  \\
 \Xi^{\ast -}    & \sqrt{1/3}  & s & s & d  \\
 \Omega^{-}      &       1/3   & s & s & s  \\
\hline \hline
\end{array}
$$
\caption{The values of $A^\prime$ and the quark flavors $q_1$, $q_2$
and $q_3$ for decuplet baryons.}
\renewcommand{\arraystretch}{1}
\addtolength{\arraycolsep}{-1.0pt}

\end{table}
Here, we would like to note  that except the $\Lambda$ current, all
octet and decuplet currents can be obtained from $\Sigma^0$ and
$\Sigma^{\ast 0}$ currents with the help of appropriate
replacements among quark flavors. In \cite{R10510}, the following relations between the
currents of the  $\Lambda$ and $\Sigma^0$ are obtained \bea
\label{e10505} 2 \eta^{\Sigma^0}(d \rar s) + \eta^{\Sigma^0} \es -
\sqrt{3}
\eta^\Lambda~, \nnb \\
2 \eta^{\Sigma^0}(u \rar s) + \eta^{\Sigma^0} \es  \sqrt{3}
\eta^\Lambda~. \eea We can now turn our attention to the calculation
of theoretical and phenomenological part of the correlation
function. In the case when pseudoscalar meson is on shell, $q^2=m_{{\cal P}}^2$, the correlation function in Eq. (\ref{e10502}) depends on two independent invariant variables, $p^2$ and $(p+q)^2$, 
the square of momenta in the two channels carried out by currents $\eta$ and $\eta_\mu$, respectively. Inserting the full set of hadrons with quantum numbers of currents $\eta$ and $\eta_\mu$ and 
isolating the ground state octet and decuplet baryons by using narrow width approximation for phenomenological part of the correlation function, we obtain:
 \bea \label{e10506}
\Pi_\mu(p,q) = { \lla 0 \vel \eta \ver {\cal O}(p_2) \rra \lla {\cal
O}(p_2) {\cal P}(q) \ve {\cal D}(p_1) \rra \lla {\cal D}(p_1) \vel
\eta_\mu \ver 0 \rra \over (p_2^2-m_{\cal O}^2) (p_1^2 - m_{\cal
D}^2) } + \cdots~, \eea where ${\cal O}(p_2)$ and ${\cal D}(p_1)$
denote the octet and decuplet baryons with momentum $p_2=p$,
$p_1=p+q$, $m_{\cal O}$ and  $m_{\cal D}$ are their masses, ${\cal
P}(q)$ is the pseudoscalar meson with momentum $q$ and $\cdots$
represents the higher states and the continuum contributions. Here, we would like to make the following remark about the Eq. (\ref{e10506}).
 Since except the $\Omega$ baryon the widths of decuplet baryons are not small, hence the narrow width approximation which have been used in the Eq. (\ref{e10506}) 
is questionable. In \cite{Erkol} it is obtained that the effect of width in calculation the mass of $\Delta$-baryon changes the result about 10\% compared to the narrow width approximation.
 For this reason, we shall neglect the width of decuplet baryons in our next discussions.

The matrix element of the interpolating current between vacuum and single
octet (decuplet) baryon state is defined in standard way
\bea
\label{e10507}
\lla 0 \vel \eta \ver {\cal O} \rra \es \lambda_{\cal O} u(p_2)~, \nnb \\
\lla {\cal D}(p_1) \vel \eta_\mu \ver 0 \rra \es \lambda_{\cal D}
\bar{u}_\mu(p_1)~, \eea where $u_\mu$ is the Rarita--Schwinger
spinor. The remaining matrix element is defined as \bea
\label{e10508} \lla  {\cal O}(p_2) {\cal P}(q) \ve {\cal D}(p_1)
\rra = g_{\cal DOP} \bar{u}(p_2) u_\mu(p_1) q^\mu~, \eea where $g$
is the coupling constant of pseudoscalar meson with octet and
decuplet baryons. Putting Eqs. (\ref{e10507}) and (\ref{e10508})
into (\ref{e10506}) and performing summation over spins of the octet
and decuplet baryons using the  formulas \bea \label{e10509}
\sum_s u(p_2,s) \bar{u}(p_2,s) \es (\rlap/{p}_2 + m_{\cal O})~, \nnb \\
\sum_s u_\mu(p_1,s) \bar{u}_\nu(p_1,s) \es -(\rlap/{p}_1 + m_{\cal
D}) \Bigg\{ g_{\mu\nu} - {\gamma_\mu \gamma_\nu \over3} - {2
p_{1\mu} p_{1\nu} \over 3 m_{\cal D}^2} + {p_{1\mu}\gamma_\nu -
p_{1\nu} \gamma_\mu \over 3 m_{\cal D}} \Bigg\}~, \eea one can
obtain the expression for the phenomenological part of the
correlation function. But, unfortunately, we face with two
drawbacks; one being that the interpolating current for decuplet
baryons does also have nonzero matrix element between vacuum and
spin--1/2 states (see \cite{R10507,R10509}), \bea \label{e10510} \lla 0 \vel \eta_\mu
\ver 1/2(p_1) \rra = (A \gamma_\mu + B p_{1\mu}) u(p_1)~, \eea
where $1/2$ stands for the spin--$1/2$ state. Multiplying both sides
of Eq. (\ref{e10510}) with $\gamma_\mu$ and using $\eta_\mu
\gamma^\mu = 0$, we get $B=-4A/m_{1/2}$. In other words, we see that
$\eta_\mu$ couples not only to spin--$3/2$, but also to unwanted
spin--$1/2$ states. From Eqs. (\ref{e10510}) and (\ref{e10506}) we
obtain that the structures proportional to $\gamma_\mu$ at the right
end and $p_{1\mu}$ contain unwanted contribution from spin--$1/2$
states, which should be removed. The second drawback in obtaining
the expression of the correlation function is related to the fact
that not all structures appearing in Eq. (\ref{e10506}) are
independent of each other. In order to cure both these problems, we
use ordering procedure of Dirac matrices as $\rlap/{q}\rlap/{p}\gamma_\mu$,. In this work, we choose
the structure $q_\mu$, which is free of the spin--$1/2$
contribution.

Using the ordering procedure, for the phenomenological part of the
correlation function we obtain: \bea \label{e10511} \Pi_\mu =
{g_{\cal DOP} (m_{\cal O} m_{\cal D}+m_{\cal D}^2-m_{\cal P}^2)\lambda_{\cal O} \lambda_{\cal D} \over
[m_{\cal D}^2-(p+q)^2] [m_{\cal O}^2-p^2]} \big\{q_\mu +
\mbox{\rm other structures}\big\}~. \eea  It follows from Eq.
(\ref{e10511}) that the interaction of pseudoscalar mesons with
decuplet--octet baryons is described by a single coupling constant.
In order to obtain the sum rule for the coupling constant $g$, the
calculation of the correlation function from QCD side is needed.
Before calculating it, we will present the relations between the
correlation functions. In other words, we try to find relations
between invariant functions for the coefficients of the structure
$q_\mu$. For establishing relations among invariant
functions, we follow the approach presented in
\cite{R10503,R10504,R10505,R10506}. The main advantage of this
approach presented below is that it takes into account $SU(3)_f$
symmetry violating effects automatically.

Similar to the works \cite{R10503,R10504,R10505,R10506}, we start
our discussion by considering the transition, $\Sigma^{\ast 0} \rar
\Sigma^0 \pi^0$. The invariant function for this transition can
formally be written in the following form \bea \label{e10512}
\Pi^{\Sigma^{\ast 0} \rar \Sigma^0\pi^0} = g_{\pi^0 \bar{u}u}
\Pi_1(u,d,s) +
 g_{\pi^0 \bar{d}d} \Pi_1^\prime(u,d,s) + g_{\pi^0 \bar{s}s} \Pi_2(u,d,s)~,
\eea
where the current for the $\pi^0$ meson is given by
\bea
\label{e10513}
\sum_{q=u,d,s} g_{\pi \bar{q}q} \bar{q} \gamma_5 q~.
\eea
For the $\pi^0$ meson we have
$g_{\pi^0 \bar{u}u} = - g_{\pi^0 \bar{d}d} = 1/\sqrt{2}$, and
$g_{\pi^0 \bar{s}s} = 0$. The invariant functions $\Pi_1$, $\Pi_1^\prime$ and $\Pi_2$
correspond to the radiation of $\pi^0$ meson from $u$, $d$ and $s$ quarks of
the $\Sigma^{\ast 0}$ baryon, respectively, and they are formally defined as
\bea
\label{e10514}
\Pi_1(u,d,s) \es \lla \bar{u}u \vel \Sigma^{\ast 0}
\Sigma^0 \ver 0 \rra~, \nnb \\
\Pi_1^\prime(u,d,s) \es \lla \bar{d}d \vel \Sigma^{\ast 0}
\Sigma^0 \ver 0 \rra~, \nnb \\
\Pi_2(u,d,s) \es \lla \bar{s}s \vel \Sigma^{\ast 0}
\Sigma^0 \ver 0 \rra~.
\eea
Since the interpolating currents of the $\Sigma^{\ast 0}$ and $\Sigma^0$
baryons are symmetric under the replacement $u \lrar d$, it is obvious that
$\Pi_1^\prime(u,d,s) = \Pi_1(d,u,s)$. Using this relation we obtain from Eq.
(\ref{e10512}) immediately that
\bea
\label{e10515}
\Pi^{\Sigma^{\ast 0} \rar \Sigma^0\pi^0} = {1\over \sqrt{2}}
[\Pi_1(u,d,s) - \Pi_1(d,u,s)]~.
\eea
Note that in the $SU(2)_f$ symmetry case, $\Pi^{\Sigma^{\ast 0} \rar
\Sigma^0\pi^0} = 0$, obviously.

The invariant function describing the $\Sigma^{\ast +} \rar
\Sigma^+\pi^0$ transition can be obtained from Eq. (\ref{e10512}) by
making the replacement $d \rar u$ and using the fact that
$\Sigma^{\ast 0} (d \rar u) = \sqrt{2} \Sigma^{\ast +}$ and
$\Sigma^0 (d \rar u) = - \sqrt{2} \Sigma^+$, which leads to the
result \bea \label{e10516} 4 \Pi_1(u,u,s) = - 2 \lla \bar{u}u \vel
\Sigma^{\ast +} \Sigma^+ \ver 0 \rra~. \eea Since $\Sigma^{\ast +}$
contains two $u$ quarks there are 4 possible ways for $\pi^0$ meson
to be radiated from the $u$ quark. Using Eq. (\ref{e10512}) and
taking into account the fact that $\Sigma^{\ast +}$ does not contain
$d$ quark, we get \bea \label{e10517} \Pi^{\Sigma^{\ast +} \rar
\Sigma^+\pi^0} = - \sqrt{2} \Pi_1(u,u,s)~. \eea The invariant
function responsible for the $\Sigma^{\ast -} \rar \Sigma^-\pi^0$
can be obtained from $\Sigma^{\ast 0} \rar \Sigma^0\pi^0$ transition
simply by making the replacement $u \rar d$ in Eq. (\ref{e10512})
and taking into account  $\Sigma^0 (u \rar d) = \sqrt{2} \Sigma^-$.
As a result we obtain \bea \label{e10518} \Pi^{\Sigma^{\ast -} \rar
\Sigma^-\pi^0} = - \sqrt{2} \Pi_1(d,d,s)~. \eea Note here that, in
$SU(2)_f$ symmetry case \bea \label{nolabel} \Pi^{\Sigma^{\ast +}
\rar \Sigma^+\pi^0} = \Pi^{\Sigma^{\ast -} \rar \Sigma^-\pi^0}~.\nnb
\eea

We now proceed by presenting the invariant functions involving $\Delta$
resonances. The invariant function for the $\Delta^+ \rar p \pi^0$
transition can be obtained from the $\Sigma^{\ast+} \Sigma^+ \pi^0$
transition by just using the identifications $\Delta^+ = \Sigma^{\ast +} (s
\rar d)$ and $p = - \Sigma^+(s \rar d)$, as a result of which we get
\bea
\label{e10519}
\Pi^{\Delta^+ \rar p \pi^0} \es - \Big[ g_{\pi^0\bar{u}u} \lla \bar{u}u \vel
\Sigma^{\ast +} \Sigma^+ \ver 0 \rra (s \rar d) +
g_{\pi^0\bar{s}s} \lla \bar{s}s \vel
\Sigma^{\ast +} \Sigma^+ \ver 0 \rra (s \rar d) \Big]\nnb \\
\es \sqrt{2} \Pi_1(u,u,d) - {1\over \sqrt{2}} \Pi_2 (u,u,d)~.
\eea
Using similar arguments, one can easily obtain the following relations
\bea
\label{e10520}
\Pi^{\Delta^0 \rar n \pi^0} \es \sqrt{2} \Pi_1(d,d,u) -
{1\over \sqrt{2}} \Pi_2 (d,d,u)~, \nnb \\
\Pi^{\Xi^{\ast 0} \rar \Xi^0 \pi^0} \es {1\over \sqrt{2}}
\Pi_2(s,s,u)~, \nnb \\
\Pi^{\Xi^{\ast -} \rar \Xi^- \pi^0} \es {1\over \sqrt{2}}
\Pi_2(s,s,d)~.
\eea
The relations presented in Eqs. (\ref{e10519}) and (\ref{e10520}) can be
further simplified by using the relation
\bea
\label{e105201}
\Pi_2(u,d,s) = - \Pi_1(s,u,d) - \Pi_1(s,d,u)~,
\eea
which we obtain from our calculations.

We can now consider the transitions involving $\eta$ meson. In this work, the
mixing between $\eta$ and $\eta^\prime$ is neglected and the interpolating
current for $\eta$ meson is chosen in the following form:
\bea
\label{e10521}
J_\eta = {1 \over \sqrt{6}} [\bar{u} \gamma_{\mu}\gamma_5 u + \bar{d}\gamma_{\mu} \gamma_5 d -
2 \bar{s} \gamma_{\mu}\gamma_5 s ]~.
\eea
In order to find relations between invariant functions involving $\eta$
meson, we choose $\Sigma^{\ast 0} \rar \Sigma^0 \eta$ as the prototype.
Similar to the $\pi^0$ case, the invariant function responsible for this
transition can be written as:
\bea
\label{e10522}
\Pi^{\Sigma^{\ast 0} \rar \Sigma^0 \eta} = g_{\eta\bar{u}u} \Pi_1(u,d,s) +
g_{\eta\bar{d}d} \Pi_1^\prime(u,d,s) + g_{\eta\bar{s}s} \Pi_2(u,d,s)~.
\eea
Using the relation given in Eq. (\ref{e105201}) we get
\bea
\label{e10524}
\Pi^{\Sigma^{\ast 0} \rar \Sigma^0 \eta} = {1 \over \sqrt{6}} [ \Pi_1(u,d,s)
+ \Pi_1(d,u,s) + 2 \Pi_1(s,u,d) + 2 \Pi_1(s,d,u)]~.
\eea

The next step in our calculation is to obtain relations between invariant
functions involving charged $\pi^\pm$ meson. Our starting point for this goal
is considering the matrix element $\lla \bar{d} d  \vel \Sigma^{\ast 0}
\Sigma^0 \ver 0 \rra $, where $d$ quarks from $\Sigma^0$ and
$\Sigma^{\ast 0}$ form the final $\bar{d}d$ state, and $u$ and $s$ are the
spectator quarks. In the matrix element $\lla \bar{u} d  \vel
\Sigma^{\ast +} \Sigma^0 \ver 0 \rra $, $d$ quark from $\Sigma^0$ and
$u$ quark from $\Sigma^{\ast 0}$ form the $\bar{u}d$ state with the
remaining $u$ and $s$ being the spectator quarks. For this reason one can
expect that these matrix elements should have relations between each other.
As a result of straightforward calculations we obtain that
\bea
\label{e10525}
\Pi^{\Sigma^{\ast 0} \rar \Sigma^+ \pi^-} \es
\lla \bar{u} d  \vel \Sigma^{\ast 0} \Sigma^+ \ver 0 \rra =
-\sqrt{2} \lla \bar{d} d  \vel \Sigma^{\ast 0} \Sigma^0 \ver 0 \rra \nnb \\
\es - \sqrt{2} \Pi_1(d,u,s)~.
\eea
Making the replacement $(u \lrar d)$ in Eq. (\ref{e10525}), we get
\bea
\label{e10526}
\Pi^{\Sigma^{\ast 0} \rar \Sigma^- \pi^+} \es
\lla \bar{d} u  \vel \Sigma^{\ast 0} \Sigma^- \ver 0 \rra =
\sqrt{2} \lla \bar{u} u  \vel \Sigma^{\ast 0} \Sigma^0 \ver 0 \rra \nnb \\
\es \sqrt{2} \Pi_1(u,d,s)~.
\eea
Following similar line of reasoning and calculations one can find the
remaining relations among the invariant functions involving charged pions,
charged and neutral $K$ mesons and $\eta$ mesons, which are presented in appendix A.

It follows from above--mentioned relations that all couplings of the
pseudoscalar mesons with decuplet--octet baryons are described with the
help of only one invariant function even if $SU(3)_f$ symmetry is violated. In other words, this approach
  takes into account the $SU(3)_f$
symmetry violation effects, automatically.
This observation constitute the principal result of the present work. 
Since the coupling constants of pseudoscalar mesons with decuplet--octet
baryons are described by only one invariant function, we need its explicit
expression in estimating their values.

As an example, we calculate the invariant function $\Pi_1$ responsible for the
$\Sigma^{\ast 0} \rar \Sigma^0 \pi^0$ transition. In deep Euclidean region, where
$-p_1^2 \rar \infty$ and $-p_2^2 \rar \infty$, the correlation function can be
calculated with the help of the operator product expansion. In obtaining 
the expression of the correlation function in LCSR from QCD side, the propagator
of light quarks, as well as the matrix elements of nonlocal operators
$\bar{q} (x_1) \Gamma q^\prime (x_2)$ and $\bar{q} (x_1) G_{\mu\nu} q^\prime
(x_2)$ between vacuum and the pseudoscalar meson are needed, where $\Gamma$
and $G_{\mu\nu}$ represents the Dirac matrices and the gluon field strength
tensor, respectively.

Up to twist--4 accuracy, the matrix elements $\lla {\cal P}(q) \vel \bar{q}(x)
\Gamma q (0) \ver 0 \rra$ and $\lla {\cal P}(q) \vel \bar{q}(x) G_{\mu\nu} q(0)
\ver 0 \rra$ are parametrized in terms of the distribution amplitudes (DA's)
as \cite{R10511,R10512,R10513}:

\bea
\label{e10527}
\lla {\cal P}(q)\vel \bar q(x) \gamma_\mu \gamma_5 q(0)\ver 0 \rra \es
-i f_{\cal P} q_\mu  \int_0^1 du  e^{i \bar u q x}
    \left( \varphi_{\cal P}(u) + {1\over 16} m_{\cal P}^2
x^2 {\Bbb{A}}(u) \right) \nnb \\
\ek {i\over 2} f_{\cal P} m_{\cal P}^2 {x_\mu\over qx}
\int_0^1 du e^{i \bar u qx} {\Bbb{B}}(u)~,\nnb \\
\lla {\cal P}(q)\vel \bar q(x) i \gamma_5 q(0)\ver 0 \rra \es
\mu_{\cal P} \int_0^1 du e^{i \bar u qx} \varphi_P(u)~,\nnb \\
\lla {\cal P}(q)\vel \bar q(x) \sigma_{\alpha \beta} \gamma_5 q(0)\ver 0 \rra \es
{i\over 6} \mu_{\cal P} \left( 1 - \widetilde{\mu}_{\cal P}^2 \right)
\left( q_\alpha x_\beta - q_\beta x_\alpha\right)
\int_0^1 du e^{i \bar u qx} \varphi_\sigma(u)~,\nnb \\
\lla {\cal P}(q)\vel \bar q(x) \sigma_{\mu \nu} \gamma_5 g_s
G_{\alpha \beta}(v x) q(0)\ver 0 \rra \es i \mu_{\cal P} \left[
q_\alpha q_\mu \left( g_{\nu \beta} - {1\over qx}(q_\nu x_\beta +
q_\beta x_\nu) \right) \right. \nnb \\
\ek q_\alpha q_\nu \left( g_{\mu \beta} -
{1\over qx}(q_\mu x_\beta + q_\beta x_\mu) \right) \nnb \\
\ek q_\beta q_\mu \left( g_{\nu \alpha} - {1\over qx}
(q_\nu x_\alpha + q_\alpha x_\nu) \right) \nnb \\
\ar q_\beta q_\nu \left. \left( g_{\mu \alpha} -
{1\over qx}(q_\mu x_\alpha + q_\alpha x_\mu) \right) \right] \nnb \\
\cp \int  D \alpha e^{i (\alpha_{\bar q} +
v \alpha_g) qx} {\cal T}(\alpha_i)~,\nnb \\
\lla {\cal P}(q)\vel \bar q(x) \gamma_\mu \gamma_5 g_s
G_{\alpha \beta} (v x) q(0)\ver 0 \rra \es q_\mu (q_\alpha x_\beta -
q_\beta x_\alpha) {1\over qx} f_{\cal P} m_{\cal P}^2
\int  D\alpha e^{i (\alpha_{\bar q} + v \alpha_g) qx}
{\cal A}_\parallel (\alpha_i) \nnb \\
\ar \left[q_\beta \left( g_{\mu \alpha} - {1\over qx}
(q_\mu x_\alpha + q_\alpha x_\mu) \right) \right. \nnb \\
\ek q_\alpha \left. \left(g_{\mu \beta}  - {1\over qx}
(q_\mu x_\beta + q_\beta x_\mu) \right) \right]
f_{\cal P} m_{\cal P}^2 \nnb \\
\cp \int D\alpha e^{i (\alpha_{\bar q} + v \alpha _g)
q x} {\cal A}_\perp(\alpha_i)~,\nnb \\
\lla {\cal P}(q)\vel \bar q(x) \gamma_\mu i g_s G_{\alpha \beta}
(v x) q(0)\ver 0 \rra \es q_\mu (q_\alpha x_\beta - q_\beta x_\alpha)
{1\over qx} f_{\cal P} m_{\cal P}^2 \int  D\alpha e^{i (\alpha_{\bar q} +
v \alpha_g) qx} {\cal V}_\parallel (\alpha_i) \nnb \\
\ar \left[q_\beta \left( g_{\mu \alpha} - {1\over qx}
(q_\mu x_\alpha + q_\alpha x_\mu) \right) \right. \nnb \\
\ek q_\alpha \left. \left(g_{\mu \beta}  - {1\over qx}
(q_\mu x_\beta + q_\beta x_\mu) \right) \right] f_{\cal P} m_{\cal P}^2 \nnb \\
    \cp \int  D\alpha e^{i (\alpha_{\bar q} +
v \alpha _g) q x} {\cal V}_\perp(\alpha_i)~.
\eea
In Eq. (\ref{e10527}) we have,
\bea
\label{nolabel}
\mu_{\cal P} = f_{\cal P} {m_{\cal P}^2\over m_{q_1} + m_{q_2}}~,~~~~~
\widetilde{\mu}_{\cal P} = {m_{q_1} + m_{q_2} \over m_{\cal P}}~, \nnb
\eea
and $D\alpha = d\alpha_{\bar q} d\alpha_q d\alpha_g
\delta(1-\alpha_{\bar q} - \alpha_q - \alpha_g)$, and
and the DA's $\varphi_{\cal P}(u),$ $\Bbb{A}(u),$ $\Bbb{B}(u),$
$\varphi_P(u),$ $\varphi_\sigma(u),$
${\cal T}(\alpha_i),$ ${\cal A}_\perp(\alpha_i),$ ${\cal A}_\parallel(\alpha_i),$
${\cal V}_\perp(\alpha_i)$ and ${\cal V}_\parallel(\alpha_i)$
are functions of definite twist whose explicit expressions can be found in
\cite{R10511,R10512,R10513}.

Propagator of the light quark in an external field is calculated in
\cite{R10514,R10515} having the form
\bea
\label{e10528}
S_q(x) \es {i \rlap/x\over 2\pi^2 x^4} - {m_q\over 4 \pi^2 x^2} -
{\lla \bar q q \rra\over 12} \left(1 - i {m_q\over 4} \rlap/x \right) -
{x^2\over 192} m_0^2 \lla \bar q q \rra  \left( 1 -
i {m_q\over 6}\rlap/x \right) \nnb \\
&&  - i g_s \int_0^1 du \left[{\rlap/x\over 16 \pi^2 x^2} G_{\mu \nu} (ux)
\sigma_{\mu \nu} - u x^\mu G_{\mu \nu} (ux) \gamma^\nu
{i\over 4 \pi^2 x^2} \right. \nnb \\
&& \left.
 - i {m_q\over 32 \pi^2} G_{\mu \nu} \sigma^{\mu
 \nu} \left( \ln \left( {-x^2 \Lambda^2\over 4} \right) +
 2 \gamma_E \right) \right]~,
\eea
where $\gamma_E \simeq 0.577$ is the Euler Constant, and following the works
\cite{R10516,R10517} $\Lambda=0.5 \div 1.0~GeV$ is used.

Using Eqs. (\ref{e10527}) and (\ref{e10528}) and choosing the coefficient
of the structure $q_\mu$, the expression of the correlation
function from QCD side can be obtained. The sum rules for the coupling
constants of the pseudoscalar mesons with decuplet--octet baryons are
obtained by matching the coefficients of the structure $q_\mu$ from
theoretical and phenomenological parts, and applying Borel transformation to
both parts with respect to the parameters $p_2^2=p^2$ and $p_1^2=(p+q)^2$
in order to suppress the higher state and continuum contributions.
As a result of these operations, we get the following sum rules for the
pseudoscalar decuplet--octet coupling constants
\bea
\label{e10529}
g_{\cal DOP} = {1\over m_{\cal O} \lambda_{\cal D} \lambda_{\cal O}}
e^{m_{\cal D}^2/M_1^2 + m_{\cal O}^2/M_2^2 + m_{\cal P}^2/(M_1^2+M_2^2)}
\Pi_{\cal DOP}~.
\eea
We only present the expression for the invariant function
$\Pi_{\cal DOP}$ for the $\Sigma^{\ast +} \rar \Sigma^+ \pi^0$ transition
in Appendix B, since invariant functions for other transitions can be
obtained from it by appropriate replacements among quark flavors. Note that for the
subtraction of continuum and higher states contribution we have used the
procedure given in \cite{R10506}.

We observe from Eq. (\ref{e10529}) that the residues $\lambda_{\cal D}$ and
$\lambda_{\cal O}$ are needed for an estimation of the coupling constants
$g_{\cal DOP}$. These residues are obtained in \cite{R10508,R10518,R10519}.
As has already been mentioned, the interpolating currents of decuplet and
octet baryons (except $\Lambda$ baryon) can be obtained from
$\Sigma^{\ast 0}$ and $\Sigma^0$ currents with the help of the corresponding
replacements. Therefore, we present the sum rules for the residues only for
$\Sigma^{\ast 0}$ and $\Sigma^0$ baryons.
\bea
\label{e10531}
\lambda_{\Sigma^0}^2 e^{-m_{\Sigma^0}^2/M^2} \es
{M^6\over 1024 \pi^2} (5 + 2 \beta + 5 \beta^2) E_2(x) - {m_0^2\over 96 M^2} (-1+\beta)^2 \lla
\bar{u} u \rra \lla \bar{d} d \rra \nnb \\
\ek {m_0^2\over 16 M^2} (-1+\beta^2) \lla \bar{s} s \rra \Big(\lla \bar{u} u \rra +
\lla \bar{d} d \rra\Big) \nnb \\
\ar {3 m_0^2\over 128} (-1+\beta^2) \Big[ m_s \Big(\lla \bar{u} u \rra +
\lla \bar{d} d \rra\Big) + (m_u+m_d) \lla \bar{s} s \rra \Big] \nnb \\
\ek  {1\over 64 \pi^2} (-1+\beta)^2 M^2 \Big( m_d \lla \bar{u} u \rra + m_u \lla \bar{d} d
\rra \Big) E_0(x) \nnb \\
\ek {3 M^2\over 64 \pi^2} (-1+\beta^2) \Big[ m_s \Big(\lla \bar{u} u \rra +
\lla \bar{d} d \rra\Big) + (m_u+m_d) \lla \bar{s} s \rra \Big] E_0(x) \nnb \\
\ar {1\over 128 \pi^2}  (5 + 2 \beta + 5 \beta^2) \Big( m_u \lla \bar{u} u \rra +
m_d \lla \bar{d} d \rra + m_s \lla \bar{s} s \rra\Big) \nnb \\
\ar {1\over 24} \Big[ 3 (-1+\beta^2) \lla \bar{s} s \rra \Big(\lla \bar{u} u
\rra + \lla \bar{d} d \rra \Big) + (-1+\beta^2) \lla \bar{u} u \rra \lla \bar{d}
d \rra \Big] \nnb \\
\ar {m_0^2\over 256 \pi^2} (-1+\beta)^2 \Big(m_u \lla \bar{d} d \rra +
m_d \lla \bar{u} u \rra\Big) \nnb \\
\ar {m_0^2\over 26 \pi^2} (-1+\beta^2) \Big[ 13 m_s \Big(\lla \bar{u} u \rra +
\lla \bar{d} d \rra\Big) + 11 (m_u+m_d) \lla \bar{s} s \rra \Big] \nnb \\
\ek {m_0^2\over 192 \pi^2}(1+\beta+\beta^2) \Big( m_u \lla \bar{u} u \rra +
m_d \lla \bar{d} d \rra - 2 m_s \lla \bar{s} s \rra\Big)\nnb \\
\ar {M^2\over 2048 \pi^4}\Big( 5+2\beta+5\beta^2\Big)E_0(x)\langle g_s^2G^2\rangle~, \nnb \\ \nnb \\
m_{\Sigma^{\ast 0}} \lambda_{\Sigma^{\ast 0}}^2 e^{-{m_{\Sigma^{\ast0}}^2\over M^2}} \es
\left( \uu + \dd + \sp \right) {M^4\over 9 \pi^2} E_1(x)
- \left( m_u + m_d + m_s\right) {M^6\over 32 \pi^4} E_2(x) \nnb \\
\ek \left( \uu + \dd + \sp \right) m_0^2 {M^2\over 18 \pi^2} E_0(x) \nnb \\
\ek {2\over 3}\left(1 + {5 m_0^2\over 72 M^2} \right) \left( m_u \dd \sp +
m_d \sp \uu + m_s \dd \uu \right) \nnb \\
\ar \left( m_s \dd \sp + m_u \dd \uu + m_d \sp \uu \right) {m_0^2\over 12
M^2}\nnb \\
\ar {5M^2\over 1152 \pi^4}E_0(x)(m_s+m_u+m_d)\langle g_s^2G^2\rangle~,
\eea
where $x = s_0/M^2$, and
\bea
\label{nolabel}
E_n(x)=1-e^{-x}\sum_{i=0}^{n}\frac{x^i}{i!}~. \nnb
\eea

\section{Numerical analysis}

In the previous section, we obtained the sum rules for the coupling constants
of the pseudoscalar mesons with decuplet--octet baryons. Here in this
section, we shall present their numerical results. In further numerical
analysis the DA's of the pseudoscalar mesons are needed, which are the main
nonperturbative parameters. These DA's and other parameters entering into
their expressions can be found in \cite{R10511,R10512,R10513}.

In the numerical analysis, $M_1^2 = M_2^2 = 2 M^2$ is chosen since
the masses of the initial and final
baryons are close to each other. With this choice, we have
$u_0 = 1/2$. The values of the remaining
parameters entering  the sum rules are:  $\langle 0|\frac{1}{\pi}\alpha_{s}G^{2}|0\rangle=(0.012
\pm 0.004)~ GeV^{4}$ \cite{Ioffeb}, $\lla \bar{u} u\rra = \lla \bar{d} d \rra = -
(0.24\pm 0.01)^3~GeV^3$, $\lla \bar{s} s\rra =0.8 \lla \bar{u} u\rra $ \cite{Ioffeb}, $m_0^2=(0.8 \pm 0.2)~GeV^2$ \cite{R10508}, $ m_{s}(2~GeV)=(111 \pm 6)~MeV$ at 
 $\Lambda_{QCD}=330~MeV$ \cite{Dominguez}, $m_u=0$, $m_d=0$, $m_\pi=0.135~GeV$, $m_\eta=0.548~GeV$, $m_K=0.498~GeV$, 
  $f_\pi=0.131~GeV$, $f_K=0.16~GeV$ and  $f_\eta=0.13~GeV$ \cite{R10511}.
Since LCSR method cannot predict the sign of $g_{\cal DOP}$,
we shall present the absolute value of it.

The sum rules for the coupling of the pseudoscalar mesons with
decuplet--octet baryons contain three auxiliary parameters, namely, Borel
mass parameter $M^2$, continuum threshold $s_0$ and the parameter $\beta$ in
the interpolating current of octet baryons. Since physical quantities should
be independent of these auxiliary parameters, it is necessary to find
regions of these parameters where the coupling constant $g_{\cal DOP}$
is independent of them.
The upper bound of $M^2$ is determined from the condition that the higher
states and continuum contributions should be less than 40--50\% of the total
value of the correlation function. The lower bound of $M^2$ is obtained by
requiring that the term with highest power in $1/M^2$ should be less than
20--25\% of the highest power of $M^2$. Using these conditions, one can easily obtain the working region for the Borel parameter $M^2$. The value of the continuum threshold
is varied in the region $2.5~GeV^2 \le s_0 \le 4.0~GeV^2$.

As an example, in Fig. (1), we present the dependence of the coupling constant
for the $\Sigma^{\ast +} \rar \Sigma^+ \pi^0$ on $M^2$ at several fixed
values of $\beta$ and at $s_0=4~GeV^2$. We see from this figure that the
coupling constant $g$ shows good stability to the variation in $M^2$, when
$M^2$ varies in the ``working region". As we already noted that the coupling constant $g_{\cal DOP}$ should be independent of auxiliary parameter $\beta$. For finding the working region of $\beta$, we present 
 the dependence of  $g_{\Sigma^{\ast +} \Sigma^+
\pi^0}$ for the $\Sigma^{\ast +} \rar \Sigma^+ \pi^0$ transition on
$\cos\theta$ as an example in Fig. (2), where $\theta$ is determined
from $\tan\theta=\beta$. We obtain from this figure that when $\cos\theta$
varies in the region $-0.5 \le \cos\theta \le 0.5$ the coupling constant
$g_{\Sigma^{\ast +} \Sigma^+ \pi^0}$
is practically independent of it. The dependence of  strong coupling constants of pion with baryons on
auxiliary parameter $\beta$ in the framework of operator product expansion at
short distance for the coorelation function of time ordering product of two currents
between the vacuum and pion states is also shown in \cite{yenidoi}. The working region of
$\beta$ in our case and  that of  given in \cite{yenidoi}  overlap but this
 is accidental.  Different problems may lead to different working regions for this auxiliary parameter.  From Fig. (2),  we see that the coupling
constant for the $\Sigma^{\ast +} \rar \Sigma^+ \pi^0$ transition is
$g_{\Sigma^{\ast +} \Sigma^+ \pi^0} = 3.4 \pm 0.5$.
The results for the coupling constants of the pseudoscalar mesons with
decuplet--octet baryons are listed in Table 3. It should be emphasized
that in this table we present only those results which cannot be obtained
from each other by simple $SU(2)_f$ rotations. The results for strong coupling constant, $g_{\cal DOP}$, when the most general form of the interpolating currents for octet baryons have been used are presented under the category "general current". In the first column of this category, the results are given in full theory.  In the second column, we present the predictions of  $SU(3)_f$ symmetry case, where $m_s=m_u=m_d=0$ and $\lla \bar{s} s\rra =\lla \bar{u} u\rra = \lla \bar{d} d \rra$. In the next category containing the columns three and four, we present our result for the  strong coupling constant, $g_{\cal DOP}$, when the Ioffe currents, $\beta=-1$ for the octet baryons have been used. The third column shows the predictions in full theory, while the presented results for the strong coupling constant in the last column have been obtained using  the $SU(3)_f$ symmetry. The errors in the presented values in Table 3 are due to the variations in Borel mass parameter, $M^2$, continuum threshold, $s_0$, auxiliary parameter $\beta$ as well as errors in input parameters entering the DA's, quark and gluon condensates, and mass of the strange quark.

A quick glance at Table 3 leads to the following conclusions.

\begin{itemize}
\item For all channels under consideration there is a good agreement between
the predictions of the general form of the current and of the Ioffe current for
the octet baryons.

\item There seems to be a considerable discrepancy between these two
predictions for the central values of $\Sigma^{*-} \rar \Lambda \pi^-$, $\Omega^{-} \rar \Xi^0K^-$, $\Delta^{0} \rar p \pi^-$, $\Sigma^{*+} \rar p \bar K^0$ and $\Sigma^{*+} \rar \Sigma^{+} \eta$ channels. 

\item Maximum value of $SU(3)_f$ symmetry  violation   is about $(10\div15 )\%$. Note that the approach presented in the present work  takes into account the $SU(3)_f$ violation effects automatically, hence we can 
 estimate order of $SU(3)_f$ violation. The essential point here is that the $SU(3)_f$ violating effects do not produce new
invariant function compared to $SU(3)_f$ symmetry case.
\end{itemize}
Finally, let us compare our predictions on coupling constants in Table 3  with the existing experimental results. Using the explicit form of the interaction Lagrangian in Eq. (\ref{e10501}), 
one can easily obtain
 expression for the decay width of the $decuplet\longrightarrow octet+pseudoscalar ~meson$ transition in terms of the strong coupling constant. 
Using the experimental values for the total widths of the $\Sigma^*$ and 
$\Xi^*$ baryons  and the branching
 ratios of $\Sigma^*\longrightarrow \Sigma \pi$,  $\Sigma^*\longrightarrow \Lambda \pi$ and  $\Xi^*\longrightarrow \Xi \pi$ \cite{pdg}, we get the following results for the related coupling constants:
\begin{equation}
g_{\Sigma^{\ast +}\Sigma^{ +}\pi^0}=3.27\pm0.55,~~~~~~ g_{\Sigma^{\ast -}\Lambda\pi^-}=4.56\pm0.48, ~~~~~~g_{\Xi^{\ast 0}\Xi^{0}\pi^0}=3.56\pm0.42
\end{equation}
Comparing these results with our predictions presented in Table 3, we see a good consistency between the values extracted from the experimental data and our predictions on the strong coupling constants 
related to the $\Sigma^{*+}\longrightarrow \Sigma^+ \pi^0$,  $\Sigma^{*-}\longrightarrow \Lambda \pi^-$ and  $\Xi^{0*}\longrightarrow \Xi^0 \pi^0$ channels. Our 
predictions on the coupling constants of channels which we have no experimental data can be verified in the future experiments.

Our concluding remarks on the present study can be summarized as follows.
The strong coupling constants of pseudoscalar mesons with
decuplet--octet baryons are investigated in LCSR by taking into account
$SU(3)_f$ symmetry breaking effects. It is seen that all coupling constants
of pseudoscalar $\pi$, $K$ and $\eta$ mesons with decuplet--octet baryons
can be represented by only one invariant function. The order of the
magnitude of $SU(3)_f$ symmetry breaking effects is also estimated.

\begin{table}[h]

\renewcommand{\arraystretch}{1.3}
\addtolength{\arraycolsep}{-0.5pt}
\small
$$
\begin{array}{|l|r@{\pm}l|r@{\pm}l|r@{\pm}l|r@{\pm}l|}
\hline \hline
 \multirow{2}{*}{$g_{\cal DOP}$}             &
 \multicolumn{4}{c|}{\mbox{General current}} &
 \multicolumn{4}{c|}{\mbox{Ioffe current}}                                                 \\
&    \multicolumn{2}{c}{\mbox{Result}}       &  \multicolumn{2}{c|}{\mbox{$SU(3)_f$}}
&    \multicolumn{2}{c}{\mbox{Result}}       &  \multicolumn{2}{c|}{\mbox{$SU(3)_f$}}      \\ \hline
 g_{_{\Sigma^{\ast +}  \Sigma^+ \pi^0}}      &  3.4&0.5 & 3.3&0.3 & 2.8&0.3 & 2.5&0.2  \\
 g_{_{\Xi^{\ast 0}  \Xi^0 \pi^0}}            &  3.3&0.7 & 3.4&0.6 & 2.4&0.2 & 2.3&0.2  \\
 g_{_{\Sigma^{\ast -}  \Lambda \pi^-}}       &  7.0&1.5 & 6.5&1.0 & 4.7&0.3 & 4.2&0.4  \\
 g_{_{\Delta^0  p \pi^-}}                    &  5.5&1.5 & 5.0&1.0 & 4.0&0.5 & 4.2&0.5  \\
 g_{_{\Delta^+  \Sigma^0 K^+}}               &  7.0&1.0 & 6.5&0.5 & 6.0&1.0 & 5.0&1.0  \\
 g_{_{\Sigma^{\ast +}  \Xi^0 K^+}}           &  3.5&0.5 & 3.6&0.4 & 3.0&0.2 & 2.8&0.2  \\
 g_{_{\Omega^-  \Xi^0 K^-}}                  &  8.0&2.0 & 7.0&1.5 & 6.5&1.0 & 6.0&1.0  \\
 g_{_{\Xi^{\ast 0}  \Sigma^+ K^-}}           &  4.7&0.6 & 4.5&0.5 & 4.8&0.8 & 4.0&0.4  \\
 g_{_{\Sigma^{\ast +}  p \bar{K}^0}}         &  6.0&1.5 & 5.0&1.0 & 4.8&0.5 & 4.4&0.4  \\
 g_{_{\Xi^{\ast 0}  \Lambda \bar{K}^0}}      &  6.4&1.0 & 5.5&1.0 & 5.0&0.6 & 4.8&0.4  \\
 g_{_{\Sigma^{\ast +}  \Sigma^+ \eta}}       &  6.0&1.0 & 5.6&1.2 & 4.8&0.4 & 4.4&0.4  \\
 g_{_{\Xi^{\ast 0}  \Xi^0 \eta}}             &  5.6&0.8 & 5.0&1.0 & 4.8&0.4 & 4.0&0.4  \\
 \hline \hline
\end{array}
$$
\caption{The values of the coupling constant $g$ for various channels.}
\renewcommand{\arraystretch}{1}
\addtolength{\arraycolsep}{-1.0pt}

\end{table}

\section*{Acknowledgments}
The authors are grateful to A. \"{O}zpineci and
V. S. Zamiralov for fruitful discussions.

\newpage

\bAPP{A}{}

In this appendix, we give the representation of correlation functions in
terms of invariant function $\Pi_1$ involving $\pi$, $K$ and $\eta$ mesons
which are not presented in the main body of the text.

\baeeq
&&\mbox{ \rm \bf{Correlation functions involving $\pi$ mesons.} }
\nnb \\ \nnb \\
\label{e105apA01}
\Pi^{\Sigma^{\ast 0} \rar \Lambda \pi^0} \es - {\sqrt{1/6}}
[2 \Pi_1(d,s,u) + \Pi_1(d,u,s) + \Pi_1(u,d,s) +
2 \Pi_1(u,s,d) ]~, \nnb \\
\Pi^{\Sigma^{\ast -} \rar \Sigma^0 \pi^-} \es \sqrt{2} \Pi_1(u,d,s)~, \nnb \\
\Pi^{\Xi^{\ast -} \rar \Xi^0 \pi^-} \es - 2 \Pi_1(d,s,s)~, \nnb \\
\Pi^{\Delta^- \rar n \pi^-} \es 2 \sqrt{3} \Pi_1(d,d,d)~, \nnb \\
\Pi^{\Sigma^{\ast -} \rar \Lambda \pi^-} \es - \sqrt{2/3}
[2 \Pi_1(u,s,d) + \Pi_1(u,d,s) ]~, \nnb \\
\Pi^{\Delta^0 \rar p \pi^-} \es 2 \Pi_1(u,u,d)~, \nnb \\
\Pi^{\Sigma^{\ast +} \rar \Sigma^0 \pi^+} \es \sqrt{2} \Pi_1(d,u,s )~, \nnb \\
\Pi^{\Sigma^{\ast 0} \rar \Sigma^- \pi^+} \es \sqrt{2} \Pi_1(u,d,s )~, \nnb \\
\Pi^{\Xi^{\ast 0} \rar \Xi^- \pi^+} \es  2 \Pi_1(u,s,s)~, \nnb \\
\Pi^{\Sigma^{\ast +} \rar \Lambda \pi^+} \es \sqrt{2/3} [ 2
\Pi_1(d,s,u ) + \Pi_1(d,u,s ) ]~, \nnb \\
\Pi^{\Delta^{++} \rar p \pi^+} \es - 2\sqrt{3} \Pi_1(u,u,u )~, \nnb \\
\Pi^{\Delta^+ \rar n \pi^+} \es - 2 \Pi_1(d,d,u)~. \nnb \\ \nnb \\ \nnb \\
&&\mbox{ \rm \bf{Correlation functions involving $K$ mesons.} }
\nnb \\ \nnb \\
\label{e105apA02}
\Pi^{\Delta^+ \rar \Sigma^0 K^+} \es - \sqrt{2} [\Pi_1(s,d,u) +
\Pi_1(s,u,d)] ~, \nnb \\
\Pi^{\Delta^+ \rar \Lambda K^+} \es \sqrt{2/3}
[\Pi_1(s,d,u) - \Pi_1(s,u,d)] ~, \nnb \\
\Pi^{\Delta^0 \rar \Sigma^- K^+} \es - 2 \Pi_1(s,d,d)~, \nnb \\
\Pi^{\Sigma^{\ast +} \rar \Xi^0 K^+} \es 2 \Pi_1(u,s,u)~, \nnb \\
\Pi^{\Sigma^{\ast 0} \rar \Xi^- K^+} \es - \sqrt{2} \Pi_1(u,s,d)~, \nnb \\
\Pi^{\Delta^{++} \rar \Sigma^+ K^+} \es 2 \sqrt{3} \Pi_1(u,u,u)~, \nnb \\
\Pi^{\Sigma^{\ast 0} \rar p K^-} \es \sqrt{2} \Pi_1(u,u,d)~, \nnb \\
\Pi^{\Omega^- \rar \Xi^0 K^-} \es - 2 \sqrt{3} \Pi_1(s,s,s)~, \nnb \\
\Pi^{\Sigma^{\ast -} \rar n K^-} \es - 2 \Pi_1(s,d,d)~, \nnb \\
\Pi^{\Xi^{\ast 0} \rar \Sigma^+ K^-} \es - 2 \Pi_1(u,u,s )~, \nnb \\
\Pi^{\Xi^{\ast -} \rar \Sigma^0 K^-} \es \sqrt{2} \Pi_1(u,d,s)~, \nnb \\
\Pi^{\Xi^{\ast -} \rar \Lambda K^-} \es - \sqrt{2/3}
[ 2 \Pi_1(u,s,d) + \Pi_1(u,d,s) ]~, \nnb \\
\Pi^{\Xi^{\ast 0} \rar \Sigma^0 \bar{K}^0} \es \sqrt{2} \Pi_1(d,u,s)~, \nnb \\
\Pi^{\Xi^{\ast 0} \rar \Lambda \bar{K}^0} \es \sqrt{2/3}
[ \Pi_1(d,s,u) + \Pi_1(d,u,s) ]~, \nnb \\
\Pi^{\Xi^{\ast -} \rar \Sigma^- \bar{K}^0} \es 2 \Pi_1(d,s,s)~, \nnb \\
\Pi^{\Sigma^{\ast 0} \rar n \bar{K}^0} \es - \sqrt{2} \Pi_1(d,d,u)~, \nnb \\
\Pi^{\Omega^- \rar \Xi^- \bar{K}^0} \es 2 \sqrt{3} \Pi_1(s,s,s)~, \nnb \\
\Pi^{\Sigma^{\ast +} \rar p \bar{K}^0} \es - 2 \Pi_1(s,u,u)~, \nnb \\
\Pi^{\Sigma^{\ast 0} \rar \Xi^0 K^0} \es \sqrt{2} \Pi_1(d,s,u)~, \nnb \\
\Pi^{\Delta^- \rar \Sigma^- K^0} \es - 2 \sqrt{3} \Pi_1(d,d,d)~, \nnb \\
\Pi^{\Sigma^{\ast -} \rar \Xi^- K^0} \es - 2 \Pi_1(d,s,d)~, \nnb \\
\Pi^{\Delta^0 \rar \Sigma^0 K^0} \es - \sqrt{2}
[ \Pi_1(s,d,u) + \Pi_1(s,u,d) ]~, \nnb \\
\Pi^{\Delta^+ \rar \Sigma^+ K^0} \es \sqrt{2}
\Pi_1(s,u,u)~. \nnb \\ \nnb \\ \nnb \\
&&\mbox{ \rm \bf{Correlation functions involving $\eta$ mesons.} }
\nnb \\ \nnb \\
\label{e105apA03}
\Pi^{\Sigma^{\ast +} \rar \Sigma^+ \eta} \es - \sqrt{2/3}
[\Pi_1(u,u,s) + 2 \Pi_1(s,u,u) ]~, \nnb \\
\Pi^{\Sigma^{\ast -} \rar \Sigma^- \eta} \es \sqrt{2/3}
[\Pi_1(d,d,s) + 2 \Pi_1(s,d,d) ]~, \nnb \\
\Pi^{\Delta^+ \rar p \eta} \es  \sqrt{2/3}
[\Pi_1(u,u,d) - \Pi_1(d,u,u) ]~, \nnb \\
\Pi^{\Delta^0 \rar n \eta} \es - \sqrt{2/3}
[\Pi_1(d,d,u) - \Pi_1(u,d,d) ]~, \nnb \\
\Pi^{\Xi^{\ast 0} \rar \Xi^0 \eta} \es - \sqrt{2/3}
[\Pi_1(u,s,s) + 2 \Pi_1(s,s,u) ]~, \nnb \\
\Pi^{\Xi^{\ast -} \rar \Xi^- \eta} \es \sqrt{2/3}
[\Pi_1(d,s,s) + 2 \Pi_1(s,s,d)]~, \nnb \\
\Pi^{\Sigma^{\ast 0} \rar \Lambda \eta} \es - (1/3\sqrt{2})
[\Pi_1(u,d,s) + 2 \Pi_1(u,s,d) + 2 \Pi_1(s,d,u) \nnb \\
\ek 2 \Pi_1(s,u,d) - 2 \Pi_1(d,s,u) - \Pi_1(d,u,s)]~. \nnb
\eaeeq

In the derivation of these results, we have used Eq. (\ref{e105201}).

\eAPP

\newpage

\bAPP{B}{}

In this appendix, we present the expression for the invariant function $\Pi$
responsible for the $\Sigma^{\ast +} \rar \Sigma^+ \pi^0$ transition,

\baeeq
\label{e105apB01}
\Pi_{\Sigma^{\ast +} \Sigma^+ \pi^0} \es
- {M^4\over 1152 \sqrt{6} \pi^2} \Big\{ 48 \mu_{\cal P} [ 4 (1+\beta) i_2({\cal T},1) - 8
\beta i_2({\cal T},v) - (1+3\beta) i_3({\cal T},1) +
2 (1+\beta) i_3({\cal T},v) ] \nnb \\
\ar 36 f_{\cal P} [ (3 + 2 \beta) m_d -
(1 - \beta) m_s ] \phi_{\cal P} (u_0) + \beta \mu_{\cal P} [6 \phi_P (u_0) +
(1 - \widetilde{\mu}_{\cal P}^2 ) (12 \phi_\sigma (u_0) -
\phi_\sigma^\prime (u_0) ) ] \Big\} \nnb \\
\ar {M^2 \over 192 \sqrt{6} \pi^2} f_{\cal P} m_{\cal P}^2 \Big\{
[(3 + 2 \beta) m_d - (1 - \beta) m_s]
[3 {\Bbb{A}}(u_0) + 32 i_2({\cal A}_\parallel,v)]
+ [m_d - (1 + \beta) m_s] \nnb \\
\cp [64 i_1({\cal A}_\parallel,1) + 64 i_1({\cal A}_\perp,1) + 3 \widetilde{i}_4({\Bbb{B}})]
+ 64 [\beta m_d - (1 + \beta) m_s] [i_1({\cal V}_\parallel,1) + i_1({\cal V}_\perp,1)] \nnb \\
\ek 128 [m_d - (1 + \beta) m_s]
[i_1({\cal A}_\parallel,v) + i_1({\cal A}_\perp,v)]
- 16 [(2 + 4 \beta) m_d + (1 - 3 \beta) m_s]
i_2({\cal V}_\parallel,1) \nnb \\
\ek 32 (1 + \beta) (2 m_d + m_s) i_2({\cal V}_\perp,1)
+ 64 (3 + 2 \beta) m_d i_2({\cal V}_\perp,v)
- 32 [(1 + \beta) m_d + \beta m_s]
i_2({\cal A}_\parallel,1) \nnb \\
\ek 64 [(1 + 2 \beta) m_d - \beta m_s]
i_2({\cal A}_\perp,1)
+ 4 {\mu_{\cal P} \over f_{\cal P}}
[(1 + \beta) i_2({\cal T},1) + 2 \beta i_2({\cal T},v)] \nnb \\
\ar {16 \pi^2 \over m_{\cal P}^2} [(3 + 2 \beta) \dd - (1 - \beta) \sp]  \phi_{\cal P}(u_0)
\Big\}\nnb \\
\ar {f_{\cal P} m_{\cal P}^2 \over 12 \sqrt{6} \pi^2}
\Bigg(\gamma_E - \ln {M^2\over \Lambda^2}\Bigg) \Big\{
\Big( m_{\cal P}^2  [(1 + \beta) m_d + \beta m_s] -
4 M^2 [(2 + \beta) m_d - m_s] \Big) \nnb \\
\cp [i_1({\cal A}_\parallel,1) + i_1({\cal A}_\perp,1)]
- \Big( m_{\cal P}^2  [(1 + \beta) m_d + \beta m_s] -
4 M^2 [(1 + 2 \beta) m_d - \beta m_s] \Big) \nnb \\
\cp [i_1({\cal V}_\parallel,1) +
i_1({\cal V}_\perp,1) ]
- \beta M^2 (m_d - 2 m_s) [2 i_2({\cal A}_\perp,1) +
i_2({\cal V}_\parallel,1) ] \nnb \\
\ar {1\over 64 M^2} [m_d - (1 + \beta) m_s]
\Big( 64 M^4  [i_2({\cal A}_\parallel,1) + 2 i_2({\cal V}_\perp,1) ] + \GG
\widetilde{i}_4({\Bbb{B}}) \Big) \nnb \\
\ek {\GG \over 32 m_{\cal P}^2} [(3 + 2 \beta) m_d -
(1 - \beta) m_s] \phi_{\cal P}(u_0) \Big\} \nnb \\
\ek { \mu_{\cal P} \GG \over 864 \sqrt{6} M^6}
(1 - \beta) (m_0^2 + 2 M^2) (1 - \widetilde{\mu}_{\cal P}^2 )
(\dd m_s + m_d \sp) \phi_\sigma(u_0) \nnb \\
\ek {f_{\cal P} \GG m_{\cal P}^2 \over 4608 \sqrt{6} \pi^2 M^2}
\Big\{[(3 + 2 \beta) m_d - (1 - \beta) m_s]
[3 {\Bbb{A}}(u_0) - 16 i_2({\cal A}_\parallel,1) + 32 i_2({\cal A}_\parallel,v)] \nnb \\
\ar 64 [m_d - (1 + \beta) m_s] [i_1({\cal A}_\parallel,1) +
i_1({\cal A}_\perp,1) - 2 i_1({\cal A}_\parallel,v) -
2 i_1({\cal A}_\perp,v)] \nnb \\
\ar 64 [\beta m_d - (1 + \beta) m_s]
[i_1({\cal V}_\parallel,1) + i_1({\cal V}_\perp,1) ]
%
%
+ 32 m_d (2 + 3 \beta) i_2({\cal A}_\perp,1) \nnb \\
\ar 32 m_d (3 + 2 \beta) [i_2({\cal V}_\perp,1) -
2 i_2({\cal V}_\perp,v)] + 16 [(2 + 3 \beta) m_d + (1 - \beta) m_s]
i_2({\cal V}_\parallel,1) \nnb \\
\ek 4 [m_d - (1 + \beta) m_s]
\widetilde{i}_4({\Bbb{B}}) \Big\} \nnb \\
\ar {\mu_{\cal P} \over 3456 \sqrt{6} M^2} \Big\{
m_0^2 (1 + 2 \beta) (\dd m_d + m_s \sp)
[6 \phi_P(u_0) - (1 - \widetilde{\mu}_{\cal P}^2 )
\phi_\sigma^\prime(u_0)] \nnb \\
\ar 96 [\beta \dd m_{\cal P}^2  m_d + 4 \dd m_0^2  m_s - 4 \beta \dd
m_0^2  m_s + m_{\cal P}^2  m_s \sp] i_2({\cal T},1) \nnb \\
%
\ar 192 m_{\cal P}^2 [\dd (1 + 2 \beta) m_d -
(1 + \beta) m_s \sp] i_2({\cal T},v) \nnb \\
\ar {6 f_{\cal P} m_{\cal P}^2 m_0^2 \over \mu_{\cal P}}
[(1 - 2 \beta) \dd - 2 (1 + 3 \beta) \sp]
\widetilde{i}_4({\Bbb{B}}) \Big\} \nnb \\
\ar {f_{\cal P} \GG \over 576 \sqrt{6} \pi^2}
[(3 + 2 \beta) m_d - (1 - \beta) m_s] \phi_{\cal P}(u_0)  -
{f_{\cal P} m_0^2 \over 288 \sqrt{6}}
[3 (5 + 4 \beta) \dd - 2 (1 - 4 \beta) \sp]
\phi_{\cal P}(u_0) \nnb \\
\ar {f_{\cal P} m_{\cal P}^4 \over 12 \sqrt{6} \pi^2}
[(1 + \beta) m_d + \beta m_s]
[i_1({\cal A}_\parallel,1) + i_1({\cal A}_\perp,1) - i_1({\cal V}_\parallel,1)
- i_1({\cal V}_\perp,1)] \nnb \\
\ek {f_{\cal P} m_{\cal P}^2 \over 144 \sqrt{6}} \Big\{
[(3 + 2 \beta) \dd - (1 - \beta) \sp]
[3 {\Bbb{A}}(u_0) - 16 i_2({\cal A}_\parallel,1) +
32 i_2({\cal A}_\parallel,v)] \nnb \\
\ek 32 \dd (2 + 3 \beta) i_2({\cal A}_\perp,1)
- 32 \dd (3 + 2 \beta) [i_2({\cal V}_\perp,1) -
2 i_2({\cal V}_\perp,v)] \nnb \\
\ek 16 [(2 + 3 \beta) \dd + (1 - \beta) \sp]
i_2({\cal V}_\parallel,1)
+ 6 [\dd - (1 + \beta) \sp] \widetilde{i}_4({\Bbb{B}}) \Big\}\nnb \\
\ar {\mu_{\cal P} \over 288 \sqrt{6}} \Big\{
\beta (1 - \widetilde{\mu}_{\cal P}^2 )
(\dd m_d + m_s \sp) \phi_\sigma^\prime(u_0)
- 16 (1 - \widetilde{\mu}_{\cal P}^2 )
(\dd m_s + m_d \sp) \phi_\sigma(u_0) \nnb \\
\ek 4 \beta  (1 - \widetilde{\mu}_{\cal P}^2 ) [\dd (m_d - 4 m_s) - (4 m_d - m_s) \sp]
\phi_\sigma(u_0)
+ 128 \dd (1 - \beta) m_s i_2({\cal T},1) \nnb \\
\ar 16 [2 \beta m_d \dd + (1 + \beta) m_s \sp] i_3({\cal T},1) -
32 (\beta m_d \dd + m_s \sp) i_3({\cal T},v) \nnb \\
\ek 6 \beta (\dd m_d + m_s \sp) \phi_P(u_0) \Big\}~, \nnb
\eaeeq
and the functions $i_n$ and $\widetilde{i}_4$  are defined as
\baeeq
\label{nolabel}
i_1(\phi,f(v)) \es \int  D\alpha_i \int_0^1 dv
\phi(\alpha_{\bar{q}},\alpha_q,\alpha_g) f(v) \theta(k-u_0)~, \nnb \\
i_2(\phi,f(v)) \es \int  D\alpha_i \int_0^1 dv
\phi(\alpha_{\bar{q}},\alpha_q,\alpha_g) f(v) \delta(k-u_0)~, \nnb \\
i_3(\phi,f(v)) \es \int  D\alpha_i \int_0^1 dv
\phi(\alpha_{\bar{q}},\alpha_q,\alpha_g) f(v) \delta^\prime(k-u_0)~, \nnb \\
\widetilde{i}_4(f(u)) \es \int_{u_0}^1 du f(u)~, \nnb
\eaeeq
where
\baeeq
k = \alpha_q + \alpha_g \bar{v}~,~~~~~u_0={M_1^2 \over M_1^2
+M_2^2}~,~~~~~M^2={M_1^2 M_2^2 \over M_1^2
+M_2^2}~.\nnb
\eaeeq

\eAPP

\newpage

\section*{Figure captions}
{\bf Fig. (1)} The dependence of the strong coupling constant
of $\pi^0$ meson with $\Sigma^{\ast +}$ and $\Sigma^+$ baryons
on Borel mass $M^2$ for several fixed values of the parameter
$\beta$ and at $s_0=4.0~GeV^2$. \\ \\
{\bf Fig. (2)} The dependence of the same coupling constant as
in Fig. (1), on $\cos\theta$ for several fixed values of the
continuum threshold $s_0$ and at $M^2=1.1~GeV^2$. \\ \\

\newpage

\begin{figure}
\vskip 3. cm
    \includegraphics{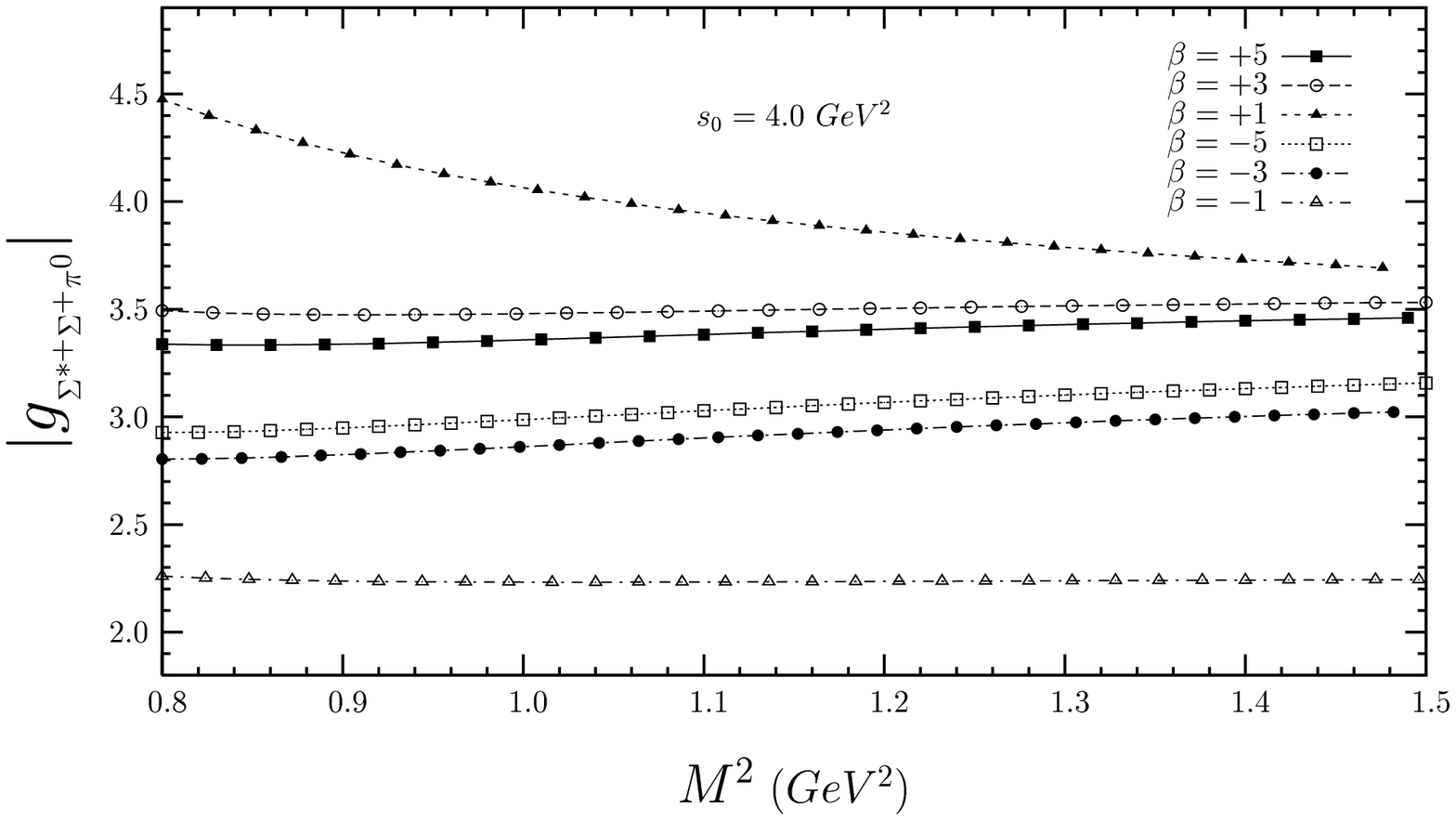}
\vskip 6.3cm
\caption{}
\end{figure}

\begin{figure}
\vskip 4.0 cm
    \includegraphics{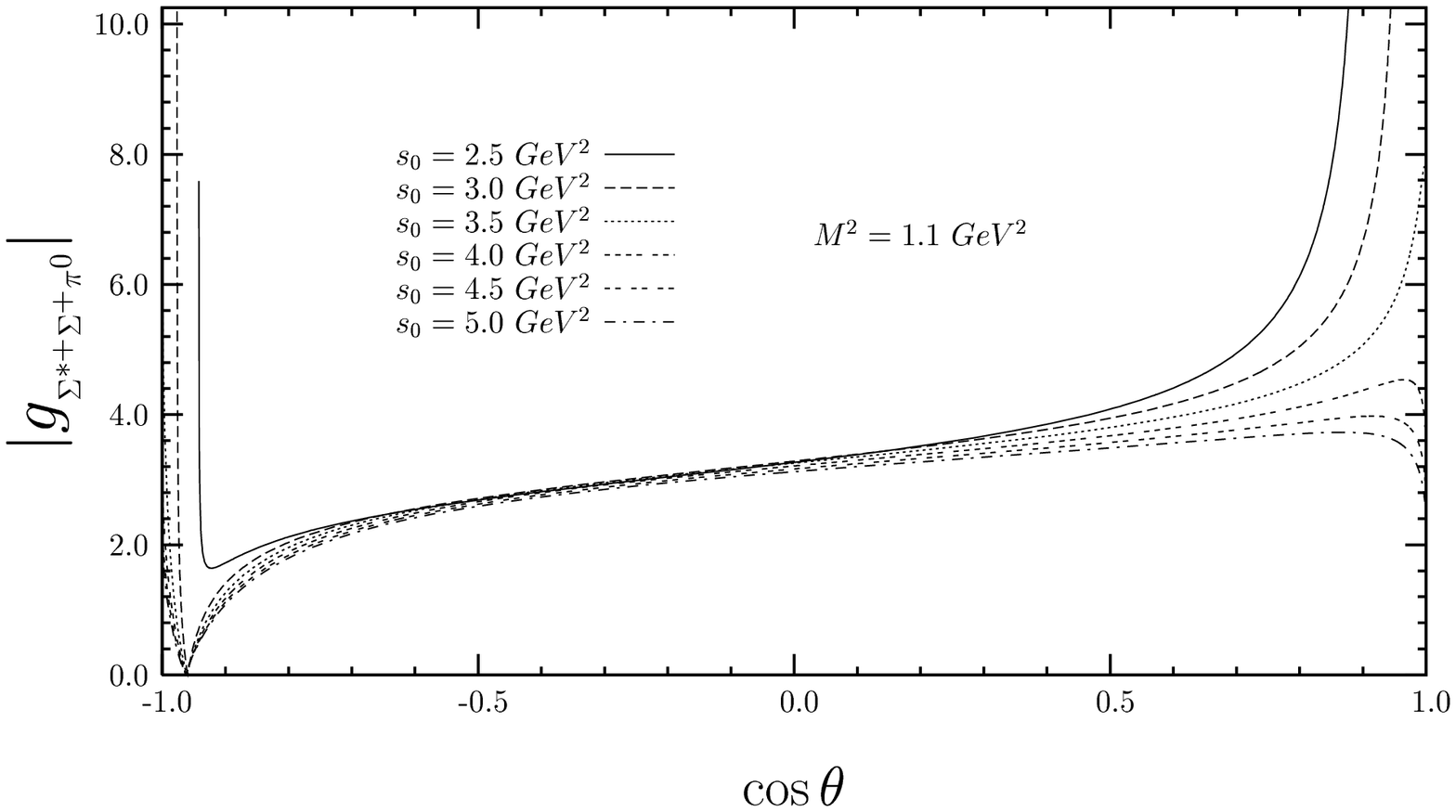}
\vskip 6.3 cm
\caption{}
\end{figure}

\end{document}